\def\XXint#1#2#3{{\setbox0=\hbox{$#1{#2#3}{\int}$}
     \vcenter{\hbox{$#2#3$}}\kern-.5\wd0}}

\documentclass[aps,prb,twocolumn,showpacs,floatfix,eqsecnum]{revtex4}

\pdfoutput=1

\usepackage{times}
\usepackage{amsfonts}
\usepackage{amssymb}
\usepackage{amsmath}
\usepackage{graphicx}
\usepackage{bm}
\usepackage{verbatim}

\usepackage{hyperref}

\usepackage{bm} 
\usepackage{color}
\usepackage{stackrel}
\usepackage{accents}

\usepackage{latexsym}

\usepackage{ulem}

\newcommand{\bsub}{\begin{subequations}}
\newcommand{\esub}{\end{subequations}}

\newcommand \bea {\begin{eqnarray} }
\newcommand \eea {\end{eqnarray}}
 
\newcommand{\beg}{\begin{equation}}
\newcommand{\en}{\end{equation}}
\newcommand{\bp}{\mathbf p}

\newcommand \bel  {\begin{align}}
\newcommand \enl  {\end{align}}

\newcommand{\lam}{\lambda}

\newcommand{\up}{\uparrow}
\newcommand{\dn}{\downarrow}
\newcommand{\dg}{^\dagger}

\newcommand{\pmat}{\begin{pmatrix}}
\newcommand{\epmat}{\end{pmatrix}}

\def\bk{{\bf k}}

\def\8{\infty}

\def\undertext#1{\vtop{\hbox{#1}\kern 1pt \hrule}}

\def\be{\begin{equation}}
\def\ee{\end{equation}}
\def\bea{\begin{eqnarray} & &}
\def\eea{\end{eqnarray}}

\def\dg{^\dagger}

\newcommand {\tcb}{\textcolor{blue}}


\begin{document}

\title{Phase diagram for the trapped $p$-wave fermionic superfluid with population imbalance}

\author{A. Kirmani, K. Quader, and M. Dzero}

\affiliation{Department of Physics, Kent State University, Kent, OH 44240 USA}

\begin{abstract} 
We consider the problem of spin-triplet $p$-wave superfluid pairing with total spin projection $m_s=0$ in atomic Fermi gas across the Feshbach resonance. We allow for imbalanced populations and take into account the effects due to presence of a parabolic trapping potential. Within the mean-field approximation for the one- and two-channel pairing models we show that depending on the distance from the center of a trap at least two superfluid states will have the lowest energy. Superfluid shells which emerge in a trap may have  two out of three angular components of the $p$-wave superfluid order parameter equal to zero.
\end{abstract}

\pacs{05.30.Fk, 03.75.Ss, 34.50.?s}

\maketitle

\section{Introduction}
The experimental discovery of the $p$-wave Feshbach resonance (FR) \cite{exp2003,exp2004Salomon,exp2004Bohn,exp2005,exp2014} 
in ultracold fermions has 
led to a flurry of theoretical work addressing various aspects of the 
superfluid pairing across the BCS-BEC crossover where the single fermionic atoms are adiabatically converted to the diatomic molecules as
one varies the resonance detuning frequency.\cite{Gurarie2007} 
Notably, interest in the physics of the $p$-wave FR in two-dimensional condensates has been rekindled due to the possibility of the topological phase transition \cite{Read2000,GurarieTopo} as well as its realization as the system is driven out-of-equilibrium.\cite{Matt1,Matt2} Cold fermion systems with unequally populated hyperfine states, and/or unequal mass, as in mixtures\cite{impmix13}, 
provide an avenue for exploration of potentially rich physics. These may also be relevant to other systems with unequal population, such as 
in quark matter\cite{quark04} and magnetic field induced superconductors.\cite{magsup2011} Past theory work on $p$-wave pairing for unequal population has been limited, and 
in the absence of a trap.\cite{imp-pwave06} 

In realistic experimental situations, however, the condensate is subject to either external optical trapping potential or to an underlying optical lattice. For the case of the $s$-wave pairing it has been shown that the presence of the optical trap together with the mass  and population imbalance leads to a variety of the superfluid states some of which are of exotic nature, such as a breached pairing state,  for example.\cite{Duan2006,Duan2006Mass,Duan2007}

\begin{figure}
\centering
\includegraphics[width=7.5cm]{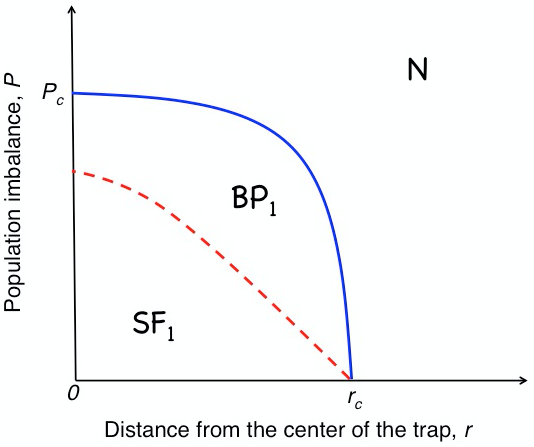}
\caption{(color online) Schematic phase diagram for the singlet $p$-wave superfluid subject to a parabolic spherically symmetric potential. State SF$_1$ denotes a doubly degenerate superfluid state which breaks time-reversal symmetry. The energy of SF$_1$ equals to the energy of TRI superfluid state SF$_0$ corresponding to the state when all angular momentum components of the pairing wave-function are nonzero. The dashed line separates the SF$_1$ state with the
superfluid breached pairing state BP$_1$ in which parts of the Fermi surface remain unpaired supporting the gapless excitations. Parameters $r_{c}$ and $P_{c}$ denote the positions in the trap and value of the population imbalance for which the first-order transition between superfluid and normal state takes place.}
\label{Fig1Main}
\end{figure}

In general, spin triplet ($s=1$, $m_s=\pm1$) $p$-wave fermion pairing with orbital quantum number $m=0,\pm 1$ can give rise to a rich variety of superfluid ground states, some of which are realized in superfluid phases of liquid $^3$He.  However, for a given system, absent additional symmetry and physical constraints, a general consideration can be daunting, even when restricted to unitary cases. As it has been discussed in Refs. [\onlinecite{Ho2005},\onlinecite{Liao2013}], considerable simplification is afforded by cold atoms subject to $p$-wave FR: different for liquid $^3$He, pairing interaction may be highly anisotropic in "spin" space ("spin" referring to hyperfine states). For example, in $^6$Li, when the hyperfine pair $|m_s, m^{\prime}_s\rangle$ = $|1/2,-1/2\rangle$ is at resonance, the pairs $|1/2, 1/2\rangle$ and  $|-1/2, -1/2\rangle$ may not be. Pairs in $p$-wave superfluids with unequal "spin" components can however have different $l=1$ components, namely $m=0,\pm 1$. Consequently, the components of the pairing wave function $\Delta_{lm}$ are related to the spherical harmonics $Y_{l}^{m}(\theta,\varphi)$.

In the context of cold atoms, spin triplet $p$-wave superfluidity with $m_s=0$ has been studied at the mean-field level, for equal population in Ref. [\onlinecite{Ho2005}], and for arbitrary population imbalance in Ref. [\onlinecite{Liao2013}], but in both cases without trapping potential. In Ref. [\onlinecite{Ho2005}], the ground state was found to be an "orbital ferromagnet",  $p_x\pm ip_y$, in which either 
\sout{$\Delta1,1$} \tcb{$\Delta_{1,1}$}  or $\Delta_{1,-1}$ paring gap component is non-zero. On the other hand, in keeping with the rotational O(3) symmetry of the system with isotropic interaction, Ref. [\onlinecite{Liao2013}] found the ground state to be degenerate with respect to the state $p_x\pm ip_y$, and the ones in which all angular momentum components of the order parameter are non-zero.
Additionally, this work suggested that, across the FR, the states which have higher energy for zero population imbalance,
e.g. the "polar" state,  $\Delta_{10}$, may acquire a lower energy for non-zero population imbalance, $P$, thereby becoming the ground state, provided $P$ exceeded some critical value. 

To the best of our knowledge the problem of the $m_s=0$ spin triplet  $p$-wave pairing with nonzero trapping potential has not been addressed yet and we attempt to do so in this paper. Specifically, we consider the pairing problem within the one- and two-channel $p$-wave pairing models and explore the effects of the spherically symmetric parabolic trapping potential by adopting the local density approximation (LDA). We utilize the mean-field theory approach to compute the ground state energy across the FR as a function of the distance from the center of the trap, $r$, and at the same time allow for arbitrary population imbalance. Our main results can be summarized on a schematic phase diagram in the $(P,r)$ plane, Fig. \ref{Fig1Main}. If we are  constrained to the center of the trap, $r=0$, and start increasing the population imbalance than there is {a phase transition between the superfluid (SF$_1$) and the normal (N) state and, in addition, within the superfluid state there exists a breached pairing state (BP$_1$) in which single fermionic excitations become gapless for some values of momenta.} Parabolic trapping potential leads to qualitatively the same physics: as the distance from the center of the trap increases the phase transition between two different superfluid states - SF$_1$ and the breached pairing state BP$_1$ -  takes place. Note, that both $r_c$ and $P_c$ are not universal and depend on the detuning frequency. For example, we find that the region of the trap where the BP$_1$ is realized gets wider as one goes from BCS to BEC side of the FR resonance.

The remaining part of the paper is organized as follows. In  Section II, we present the two-channel model, basic equations and approximations we use to study the $m_s =0$ triplet $p$-wave superfluid in a parabolic trap. 
We also present a Ginzburg-Landau analysis which we later use as a guide to obtain solutions of the mean-field self-consistency equations. In Section III we present and discuss our main results for the two-channel model. 
We also present results and discussions of the single-channel model, relevant for wide resonances in Section IV. 
Sections V  and VI  contain concluding discussion and acknowledgments. 
Lastly,  in Appendix A we provide detailed analysis of the self-consistency equations which guide the search for the system's ground state energy as a function of the distance from the center of the trap. 

\section{Two-channel model and basic equations}
We present a two-channel model which describes fermionic atoms in two hyperfine states interacting via $p$-wave $m_s=0$ triplet Feshbach resonance. We consider systems with arbitrary population imbalances. 
In our two-channel model  molecules with non-zero center-of-mass momenta are ignored. With this provision the model Hamiltonian reads
\beg\label{Eq1}
\begin{split}
H=&\sum\limits_{\bk\sigma}\epsilon_{\bk\sigma}\hat{c}_{\bk\sigma}\dg \hat{c}_{\bk\sigma}+\omega\sum\limits_{m=-1}^{1}\hat{b}_m\dg \hat{b}_m\\&+g\sum\limits_{\bk}\sum\limits_{m=-1}^1w_\bk\left[{Y}_{1m}(\hat{\bk})\hat{b}_m\hat{c}_{\bk\up}\dg\hat{c}_{-\bk\dn}\dg+\textrm{h.c.}\right]
\end{split}
\en
Here $g$ is a coupling constant, $\hat{c}_{\bk\sigma}\dg, \hat{c}_{\bk\sigma}$ are creation and annihilation fermionic operators, 
$\sigma=\up,\dn$ denotes the two hyperfine states and $\epsilon_{\bk\sigma}$ are single particle energies are given by
\beg\label{DefineH}
\epsilon_{\bk\sigma}=\frac{k^2}{2m}-h\cdot\textrm{sign}(\sigma),
\en 
$\hat{b}_m\dg, \hat{b}_m$ are the bosonic operators which describe the creation and annihilation of molecules with the orbital quantum number 
$m=0,\pm 1$ of binding energy $\omega$. The parameter $h$ accounts for the population imbalance between the two hyperfine states and function $w_\bk=k_0k/(k^2+k_0^2)$ with $k_0\sim k_F$ guarantees the convergence of the momenta summations, so that one does not need
to introduce the ultraviolet cutoff. Just as in the case of an $s$-wave condensate, the model (\ref{Eq1}) describes superfluid fermions - BCS side of the Feshbach resonance - when $\omega$ exceeds the Fermi energy $\varepsilon_F$ and bound molecules when $\omega$ is decreased below the Fermi energy, so that deep in the BEC regime $\omega<0$ and $|\omega|\gg\varepsilon_F$. 

In the mean-field approximation, the bosonic operators in the Hamiltonian (\ref{Eq1}) are replaced with their expectation values, 
$\hat{b}_m\to b_m=\langle\hat{b}_m\rangle$. In what follows, it will be convenient to introduce the pairing fields
\beg\label{Delta}
\Delta_m=-gb_m.
\en
In complete analogy with the discussion on the mean-field theory for the the two-channel $s$-wave model  in Ref. \onlinecite{BigQuench2015}, 
we obtain the following zero-temperature self-consistency equations for the pairing field components
\beg\label{SelfConsist}
\begin{split}
\Delta_m=&\frac{g^2}{(\omega-2\mu)}\sum\limits_{\bk}\frac{w_k^2\theta(E_\bk-h)}{2\sqrt{\xi_\bk^2+|\Delta(\bk)|^2}}\\&\times\sum\limits_{n=-1}^{1}{Y}_{1m}^*(\hat{\bk})Y_{1n}(\hat{\bk})\Delta_n,
\end{split}
\en
where $\xi_\bk=k^2/2m-\mu$, $E_\bk=\sqrt{\xi_\bk^2+|\Delta(\bk)|^2}$, superfluid order parameter 
$\Delta(\bk)=w_k\sum\limits_mY_{1m}(\hat{\bk})\Delta_m$ and 
$\mu$ is the chemical potential determined from the particle number equation
\beg\label{PartNumber}
2n=\frac{2}{g^2}\sum\limits_{m=-1}^1|\Delta_m|^2+\sum\limits_{\bk}\left[1-\frac{\xi_\bk}{E_\bk}\theta(E_\bk-h)\right],
\en
where $n=(n_\up+n_\dn)/2$ is a total particle density. In Appendix A we provide the detailed analysis of the possible roots of 
Eqs. (\ref{SelfConsist}) which will help us with the analysis of the case of the non-zero trapping potential.

In analogy with the $s$-wave case \cite{Gurarie2007,BigQuench2015}, we introduce the dimensionless parameter 
\beg\label{gamma}
\gamma=\frac{g^2\nu_F}{\varepsilon_F}
\en 
and $\nu_F$ is the density of states at the Fermi level. This parameter describes the width of the Feshbach resonance and has a physical meaning of the dimensionless interaction strength between 
the atoms and molecules. For broad Feshbach resonance, $\gamma\gg 1$, the singlet $p$-wave pairing problem can be 
addressed in terms of the following single-channel model:\cite{Ho2005,Liao2013}
\beg\label{BCS}
H_{\textrm{1ch}}=\sum\limits_{\bk\sigma}\xi_{\bk\sigma}\overline{c}_{\bk\sigma} c_{\bk\sigma}-\sum\limits_{\bk\bp}
V_{\bk\bp}
\overline{c}_{\bk\up}\overline{c}_{-\bk\dn} c_{-\bp\dn}c_{\bp\up},
\en
where $V_{\bk\bp}=(\lambda/\nu_F) w_\bk w_\bp\sum_{mn}Y_{1m}^*(\hat{\bk})Y_{1n}(\hat{\bp})$ and $\lambda>0$ is the dimensionless pairing strength. It is important to 
note that while the mean-field approximation for the model (\ref{BCS}) holds deep in the BEC regime corresponding to $\lambda\gg 1$ as well
as in the BCS regime $\lambda\ll 1$, it is not applicable for the intermediate values of the coupling constant, $\lambda\sim1$. That is why in this paper we will focus more on the physics governed by the 
two-channel model (\ref{Eq1}), though we also present our results for the one-channel model in Section IV.
\begin{figure}
\centering
\includegraphics[width=8cm]{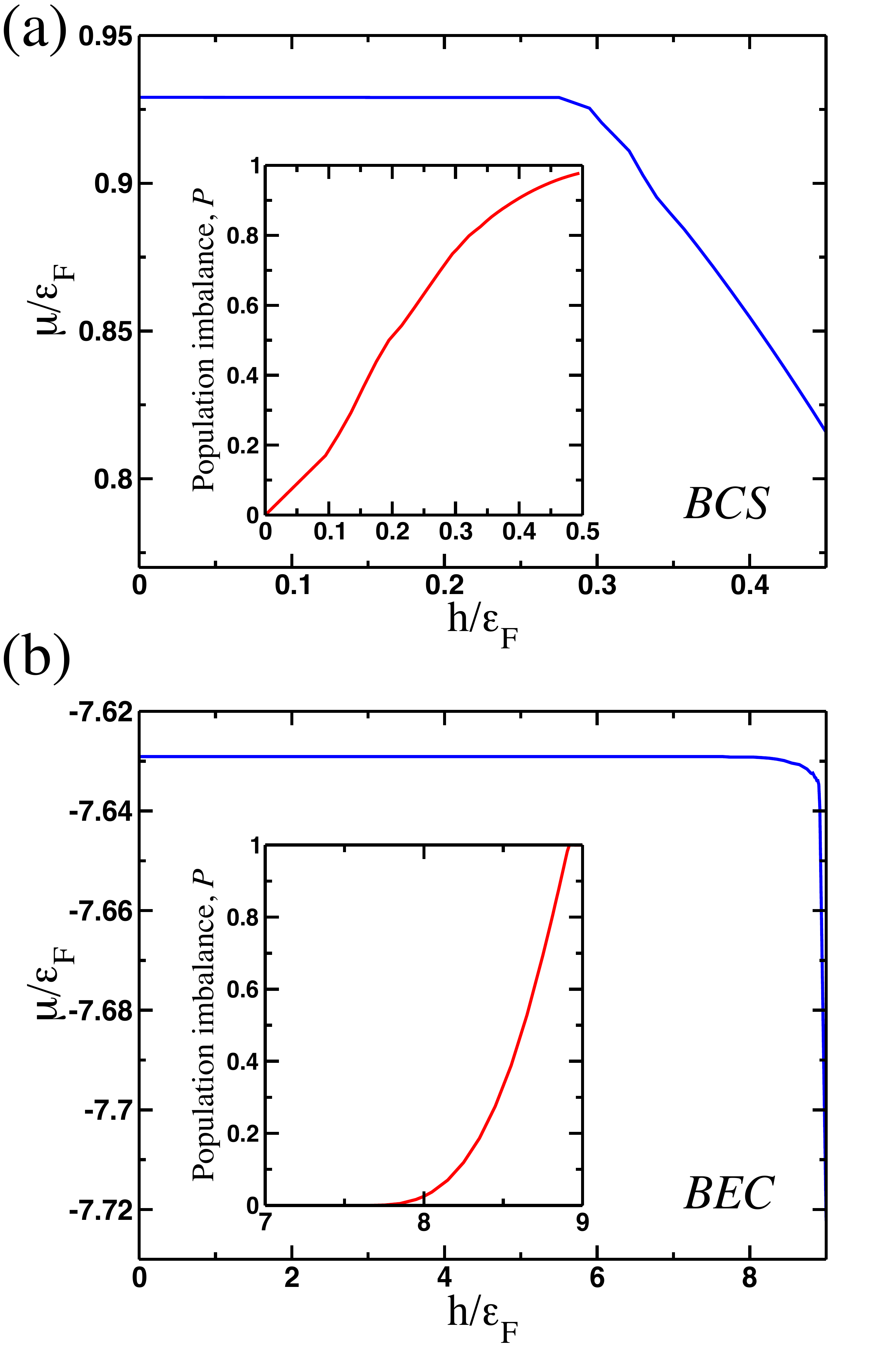}
\caption{Chemical potential and population imbalance as a function of parameter $h$ (in the units of Fermi energy $\varepsilon_F$) calculated by solving the system of self-consistency and the particle number equations for a parabolic trapping potential. The results are shown for the following choice of parameters: 
$\gamma=1.45$, $n=0.875$, while the detuning frequency is $\omega=2\mu_0+g^2\nu_F/3.15$ (top panel) and $\omega=2\mu_0+g^2\nu_F/5.96$ (inset), where $\mu_0$ is a chemical potential for $h=0$ (bottom panel).}
\label{Fig2ChemPot}
\end{figure}

\begin{figure}
\centering
\includegraphics[width=8cm]{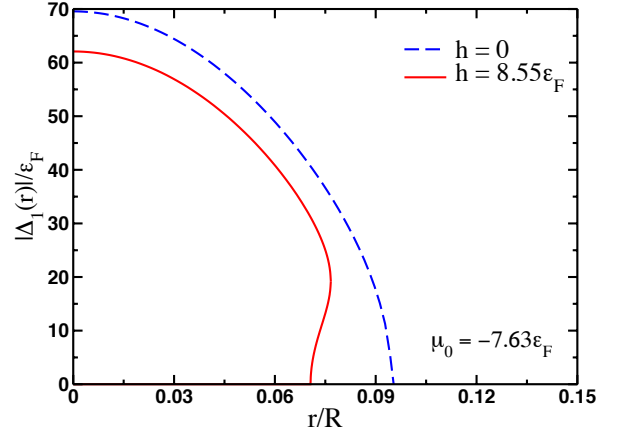}
\caption{Radial dependence of the pairing amplitude $|\Delta_1(r)|$ (SF$_1$) for zero and finite values of the parameter $h$ (in the units of Fermi energy $\varepsilon_F$) calculated by solving the system of self-consistency and the particle number equations for a parabolic trapping potential. The results are shown for the following choice of parameters: 
$\gamma=1.45$, $n=0.875$, while the detuning frequency is $\omega=2\mu_0+g^2\nu_F/5.96$.}
\label{Fig3DLT1r}
\end{figure}

\subsection{Ginzburg-Landau expansion}
To get insight into the energy structure of the $p$-wave pairing state, it will be useful to utilize the Ginzburg-Landau approach ignoring the trapping potential. At temperatures just below the critical temperature $|T-T_c|/T_c\ll 1$ all order parameter components are small and we can expand the right hand side of the self-consistency equation in powers of $\Delta_m=|\Delta_m|e^{i\phi_m}$ ($m=0,\pm1$). The free energy which corresponds to this expansion has the following form
\beg\label{FreeEnergy}
\begin{split}
{\cal F}[\Delta]=&{\cal F}_0-{a}(h,T)\sum\limits_{m=-1}^{1}|\Delta_{m}|^2\\&+b(h,T)\sum\limits_{kl}\sum\limits_{mn}{\cal I}_{mn}^{kl}
\Delta_{m}^*{\Delta}_{n}^*\Delta_{k}\Delta_{l},
\end{split}
\en
where ${\cal F}_0$ is the free energy in the normal state, the expansion coefficients $a(h,T)>0$ and $b(h,T)>0$ are some known functions of the population imbalance and temperature.  
Using the symmetry of the fourth term with respect to permutations $m\leftrightarrow n$ and $k\leftrightarrow l$, we can compactly rewrite the expression for the coefficients ${\cal I}_{mn}^{kl}$ as follows
\beg\nonumber
\begin{split}
{\cal I}_{mn}^{kl}&=-2\left[\delta_{m,0}\delta_{n,0}\delta_{k,-1}\delta_{l,1}+\delta_{m,-1}\delta_{n,1}\delta_{k,0}\delta_{l,0}\right]\\&+4
\left[\delta_{m,-1}\delta_{n,0}\delta_{k,-1}\delta_{l,0}
+\delta_{m0}\delta_{n,1}\delta_{k,1}\delta_{l0}
\right]\\&+2\left[\delta_{m,-1}\delta_{n,-1}\delta_{k,-1}\delta_{l,-1}+\delta_{m,1}\delta_{n,1}\delta_{k,1}\delta_{l,1}\right.
\\&\left.+4\delta_{m,-1}\delta_{n,1}\delta_{k,1}\delta_{l,-1}\right]+3\delta_{m,0}\delta_{n,0}\delta_{k,0}\delta_{l,0}.
\end{split}
\en
A quick analysis of the free energy (\ref{FreeEnergy}) shows that the conditions for the minima of the free energy are satisfied for 
\beg\label{phases}
2\phi_0-\phi_{1}-\phi_{-1}=2\pi n, ~(n=0,1,2,...)
\en
Let us introduce the following three-component vector ${\vec \Delta}=(|\Delta_{-1}|,\Delta_0,|{\Delta}_1|)$ 
and we take into account that $\Delta_0$ can be considered purely real due to condition (\ref{phases}). 
The free energy has a minimum ${\cal F}_{\textrm{min}}=-{50 a^2}/{9b}$ corresponding to a two superfluid states which differ from each other by symmetry. One of these two states denoted by SF$_0$ is described by ${\vec \Delta}$ with all nonzero components which satisfy 
\beg\label{globalrels}
|\Delta_{-1}|=|\Delta_{1}|, \quad \Delta_0=\sqrt{2}|\Delta_1|.
\en 
The other state denoted by SF$_1$ is described by ${\vec \Delta}_{\textrm{fm}}=(\Delta_{-1},0,0)$ or 
${\vec \Delta}_{\textrm{fm}}=(0,0,\Delta_{1})$: it breaks time-reversal symmetry, it is doubly degenerate and given the spin-triplet nature of the pairing it corresponds to an orbital ferromagnet. \cite{Ho2005,Liao2013} It can be easily checked that the length of the vector 
${\vec \Delta}$ is the same in both of these states. Furthermore, inclusion of the sixth order terms in the Ginzburg-Landau expansion lead to the same result: the states SF$_0$ and SF$_1$ will have the same free energy below $T_c$ since the system gains exactly the same amount of energy condensing into one of these states. Also note that the accidental degeneracy between these two states implies that the corresponding chemical potentials will also be the same.  

The multicomponent nature of the spin-triplet ($m_s=0$) $p$-wave superfluid furnishes another possible solutions which have somewhat higher energy than ${\cal F}_{\textrm{min}}$. {For} our subsequent discussion it is useful to briefly mention these states here. There are two particular order parameter configurations corresponding to the partially condensed states: the first state which we denote as 
{SP$_2$} is defined by $(\Delta_0=\sqrt{2/3}|\Delta_\textrm{fm}|, \Delta_{1}=\Delta_{-1}^*=0)$ while the second one SP$_3$  corresponds 
to $(\Delta_0=0,\Delta_{\pm 1}\not=0)$ with $|\Delta_{\pm 1}|=\sqrt{2/3}|\Delta_{\textrm{fm}}|$. As it turns out,
both of these states at $T\sim T_c$ have the same energy ${\cal F}_{\textrm{opm}}=-100 a^2/27b=2{\cal F}_{\textrm{min}}/3$.  In what follows, we will use the insights from the spatially homogeneous Ginzburg-Landau theory to analyze the ground state properties in the presence of the parabolic trapping potential. We will be mainly interested in finding out whether the ground state energy configuration changes with distance from the center of an optical trap. In particular, 
the phase relation (\ref{phases}) becomes useful in identifying the order parameter configurations with the lowest energy. 

\subsection{Self-consistency equations in the local density approximation}
We include the effects of the trapping potential $V({\mathbf r})$ using the local density approximation (LDA). At the heart of the LDA approach is an assumption that the physical quantities do not change substantially on the length scale of the trapping potential. Thus, the LDA is valid when the 
size of the pairing gap greatly exceeds the single particle level spacing at the Fermi level.\cite{Pit2008} Within the LDA scheme, the gradient terms of the density and pairing amplitude are neglected, so that one can adopt the Thomas-Fermi theory to describe the superfluid pairing. In principle, the 
corrections to the ground states due to the gradient terms can be found by perturbation theory. \cite{CorrectLDA}

We consider the spherically symmetric trapping potential $V({\mathbf r})=\beta r^2/2$.  Formally, the LDA is implemented by considering the nonlocal chemical potential 
\beg\label{ChemLDA}
\mu({\mathbf r})=\mu-V({\mathbf r}).
\en 
One could define center-potential to be $\mu_0$. Then, the Fermi energy is related to the particle number by 
$\varepsilon_F=(3N\beta^{3/2})^{1/3}/{\sqrt{m}}$ and the particles occupy the spherical volume of radius $R=\sqrt{2\varepsilon_F/\beta}$. Given (\ref{ChemLDA}) it follows that the order parameter components (\ref{SelfConsist}) and the particle density (\ref{PartNumber}) depend on the position relative to the center of the trap, $\Delta_m\to\Delta_{m}({\mathbf r})$, $n\to n({\mathbf r})$. 
The total particle number in a trap is fixed, {so} in order to determine the global chemical potential $\mu$, Eq. (\ref{ChemLDA}), we need to integrate both parts of the equation for $n({\mathbf r})$ over the trap and normalize the resulting integrals by the volume of the trap. Since $n({\mathbf r})$ depends on $\Delta_m({\mathbf r})$, at each step of the calculation the self-consistency equations for $\Delta_m({\mathbf r})$ must
be solved. Note, that unlike the case of the single channel model, the effective coupling constant $\lambda_{\textrm{eff}}({\mathbf r})=g^2/(\omega-2\mu({\mathbf r}))$ becomes dependent on the distance from the center of the trap and the population imbalance. 

After the global chemical potential {is} found, we determine the spatial profile of the order parameter components $\Delta_m({\mathbf r})$ allowing for all possible configurations as we have discussed above. At zero temperature, the ground state configuration at each point of the
trap can be identified by computing the local energy
\beg\label{LocalEnergy}
\begin{split}
E_{\textrm{gs}}=&\sum\limits_{\bk}\left[\xi_\bk-E_\bk+(E_\bk-h)\cdot\theta(h-E_\bk)\right]\\&+
\sum\limits_{m=-1}^{1}\frac{(\omega-2\mu)}{g^2}|\Delta_m|^2,
\end{split}
\en 
where we have omitted the dependence on ${\mathbf r}$ for brevity. Lastly, we note that to describe the population imbalance we will work with the parameter $h$ rather {than} the parameter $P=(N_{\up}-N_{\dn})/(N_\up+N_\dn)$ for convenience. However, we will quote both values $(h,P)$ where necessary. 

\section{Phase diagram for two-channel model}

In this Section we summarize our results for the two-channel case. But, first, we provide few comments on the procedure we have used. In order to compute the dependence of the pairing amplitude on the distance from the center of a trap, we need to compute the global chemical potential from the particle number equation $N=\int d^3{\mathbf r}[n_\up({\mathbf r})+n_{\dn}({\mathbf r})]$ normalized by the volume of the trap and the expression under the integral is given by the right hand side of the equation (\ref{PartNumber}) while the integration should be performed over the trap volume. After the global chemical potential has been determined we solve the self-consistency equations (\ref{SelfConsist}) taking into account (\ref{ChemLDA}) and the dependence of the effective coupling on ${\mathbf r}$. Note that due to the fact that the chemical 
potential remains the same across the trap guarantees that the only superfluid state which is realized in the trap in the SF$_1$ state since it will always have the lowest energy among the all possible pairing states. This is in sharp contrast with the case of no trapping potential when for a given population imbalance, each pairing state has its own chemical potential 
and the ground state is determined by comparing the energies of the corresponding superfluid states. 

{In} Fig. \ref{Fig2ChemPot} we show the results for the dependence of the global chemical potential $\mu$, Eq. (\ref{ChemLDA}),  {on} the parameter $h$, Eq. (\ref{DefineH}), for the superfluid states SF$_1$. As we have mentioned above, the non-trivial feature of the two-channel model is that the effective pairing strength depends on both population imbalance and trapping potential via its dependence on the chemical potential. Moreover, the difference in the values of the global chemical potential appears to be the main reason for the deviations from the ground and metastable energy configurations predicted by the Ginzburg-Landau expansion. 

\paragraph{BEC regime.} In Fig. \ref{Fig3DLT1r} we show the radial dependence of the pairing amplitudes on far BEC side of the FR. As one can see, at higher values of the population imbalance the pairing amplitude shows nonlinear dependence on $r$. This non-linearity signals an instability of the spatially homogeneous solution at given $r$. {Specifically, it suggests that if the size of the optical trap is large enough, the Cooper pairs will acquire some finite center-of-mass momentum forming a spatially inhomogeneous superfluid since the energy of this state becomes lower than the energy of the spatially homogeneous one. In conventional superconductors this state has been dubbed in literature as Fulde-Ferrel-Larkin-Ovchinnikov (FFLO) state. }

The phase diagram in the BEC regime is shown on Fig. 4(a). At low values of the population imbalance and close to the center of the trap the superfluid remains unpolarized and the single particle excitation spectrum remains gapped for all values of momenta. As one increases the value of the population imbalance, the system develops local polarization, 
$n_\up({\mathbf r})-n_{\dn}({\mathbf r})\not=0$, which also corresponds to the emergence of the gapless excitations in given momentum interval where $\sqrt{\xi_\bk^2+|\Delta(\hat{\bk})|^2}-h\leq 0$. We call this state a breached paired state (BP$_1$) \cite{DuanBP1,DuanBP2} and the dashed line separates unpolarized SF$_1$ and BP$_1$ states. 

\begin{figure}
\centering
\includegraphics[width=7.5cm]{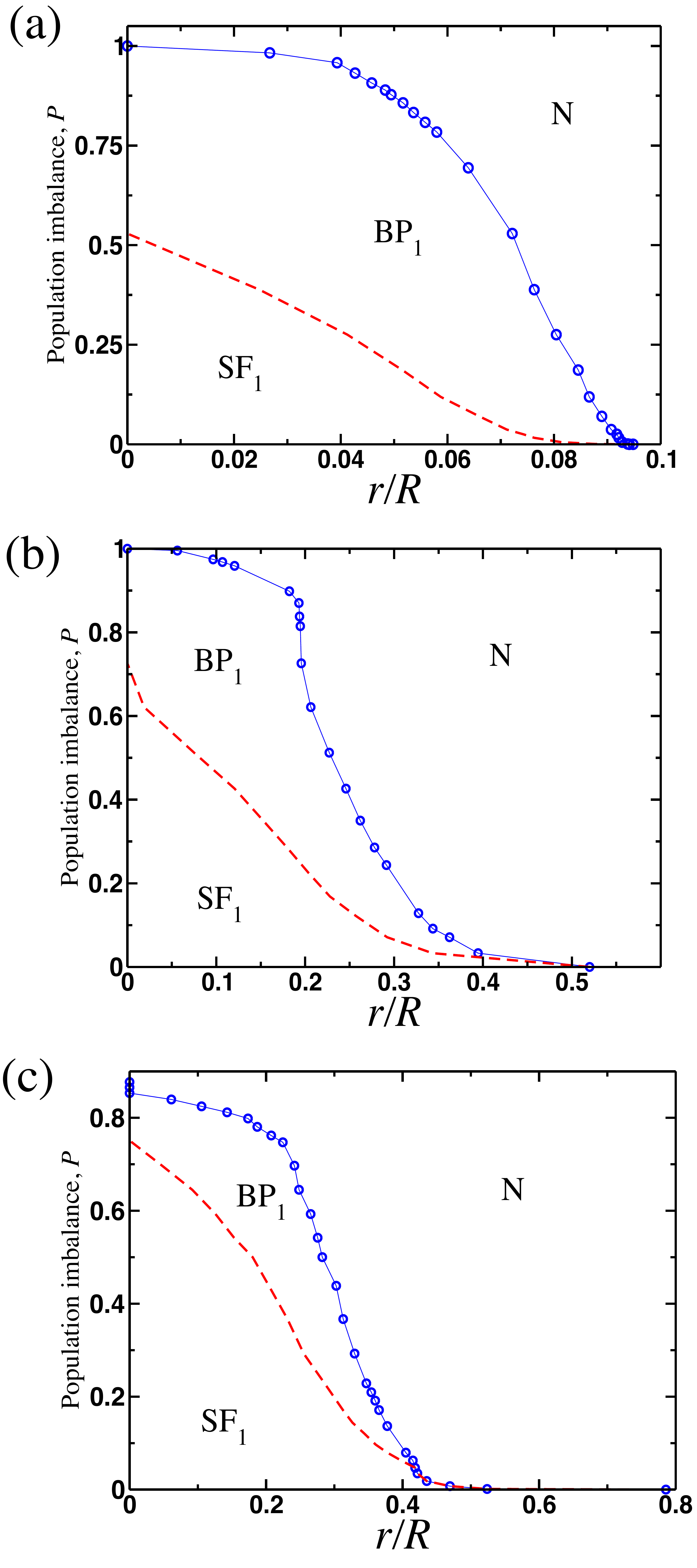}
\caption{Phase diagram for the BEC (top panel), crossover (middle panel) and BCS (bottom panel) regions in the $(r,P)$ plane extracted from the data for the ground state energy density as a function of the distance from the center of the trap. The detuning frequencies are $\omega=2\mu_0+g^2\nu_F/2.35$ for BCS region, $\omega=2\mu_0+g^2\nu_F/3.17$ for the crossover region and $\omega=2\mu_0+g^2\nu_F/5.95$ for the BEC regions of the FR.}
\label{Fig4PD}
\end{figure}

\paragraph{Crossover and BCS regimes.} 
We carried out the calculation similar to the one above for the case {when the} detuning frequency is at close to the crossover regime so that the corresponding chemical potential is close to zero. We show the plot of the phase diagram on Fig. 4(b). The only quantitative difference with the BEC case is the narrowing of the region of the breach-pairing state. 
Lastly, the phase digram in the BCS regime, Fig. 4(c) is again qualitatively similar to the diagram for two other regimes. Lastly, we note that unlike BEC case, in BCS and crossover regimes, the superfluid state extends significantly far from the center of the trap. 

{\section{Phase diagram for one-channel model}}

In this Section we present summary of the results for one-channel model. As we have emphasized {above} there are two main difference here compared to the two-channel model: (1) the coupling constant remains independent {of} the position in the trap and (2) the particle number equation does not have the term proportional to $|\Delta|^2/\lambda$ i.e. the first term on the right-hand-side of equation (\ref{PartNumber}). In one-channel model the self-consistency and the number equations take the following form

\beg\label{SelfConsist1ch}
\begin{split}
	\Delta_m&=\lambda \sum\limits_{E_\bk\geq h}\frac{w_k^2}{2E_\bk}\sum\limits_{n=-1}^{1}{Y}_{1m}^*(\hat{\bk})Y_{1n}(\hat{\bk})\Delta_n,\\
2n&=\sum\limits_{\bk}\left[1-\frac{\xi_\bk}{E_\bk}\theta(E_\bk-h)\right],
\end{split}
\en
Just as in case of two-channel model we have $n=(n_{\up}-n_{\dn})/2$ and the gap amplitudes and particle density are functions of trap coordinates. We can also identically obtain the above equation by using method of Green's function and deploying Matsubara's summations. In Local Density Approximation (LDA), it is easy to show that the total particle number can be re-written as
\begin{figure}
	\centering
	\includegraphics[width=7.5cm]{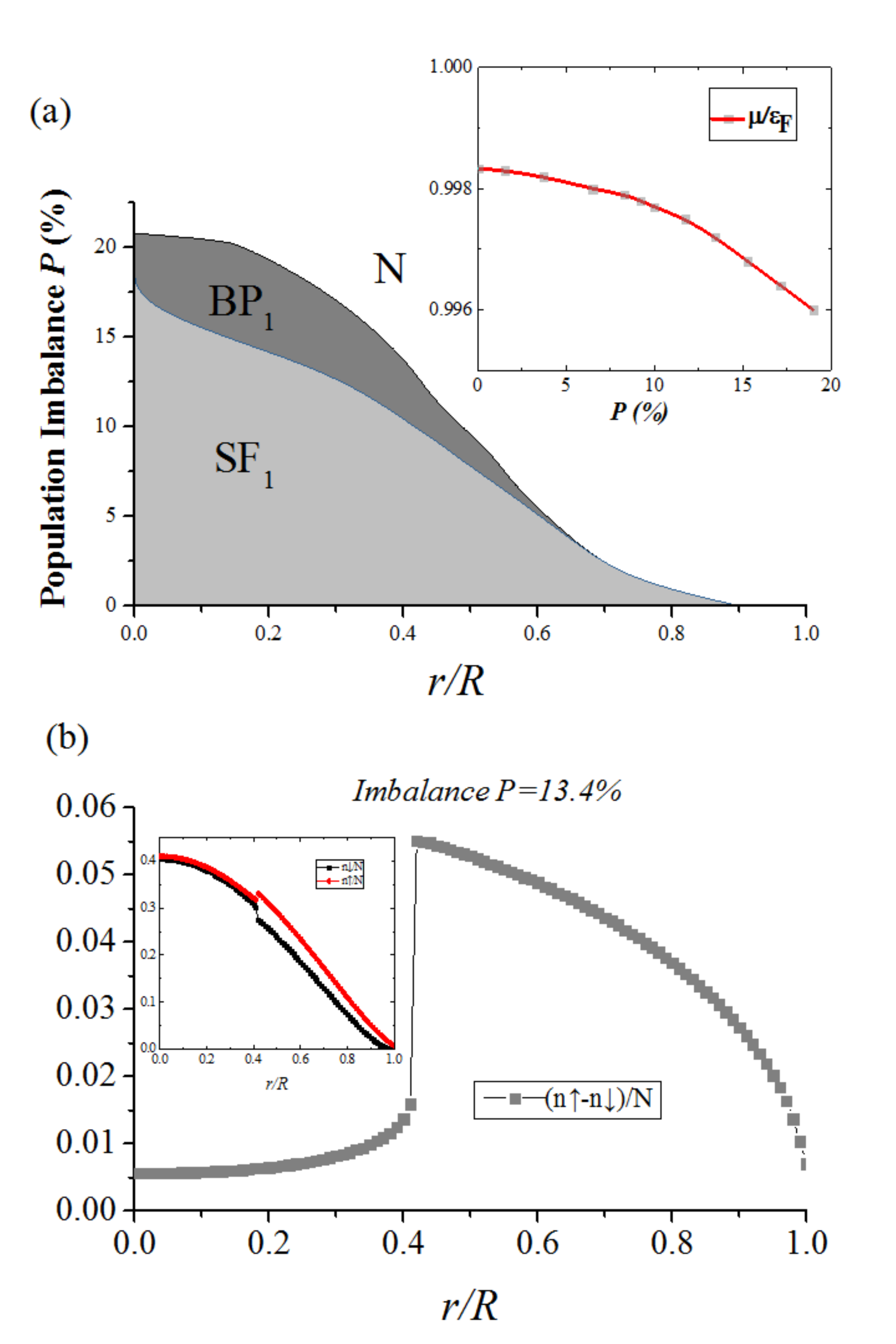}
	\caption{ {\bf  (a)} Phase diagram for the BCS region in the $(r,p)$ plane extracted from the data for the ground state energy density as a function of the distance from the center of the trap. Inset is $\mu_0$ as function of Imbalance. {\bf  (b)} local spin/population densities and its ratio with local total number. The coupling constant is $\lam=1.45$.}
	\label{Fig5}
\end{figure}

\begin{figure}
	\centering
	\includegraphics[width=7.5cm]{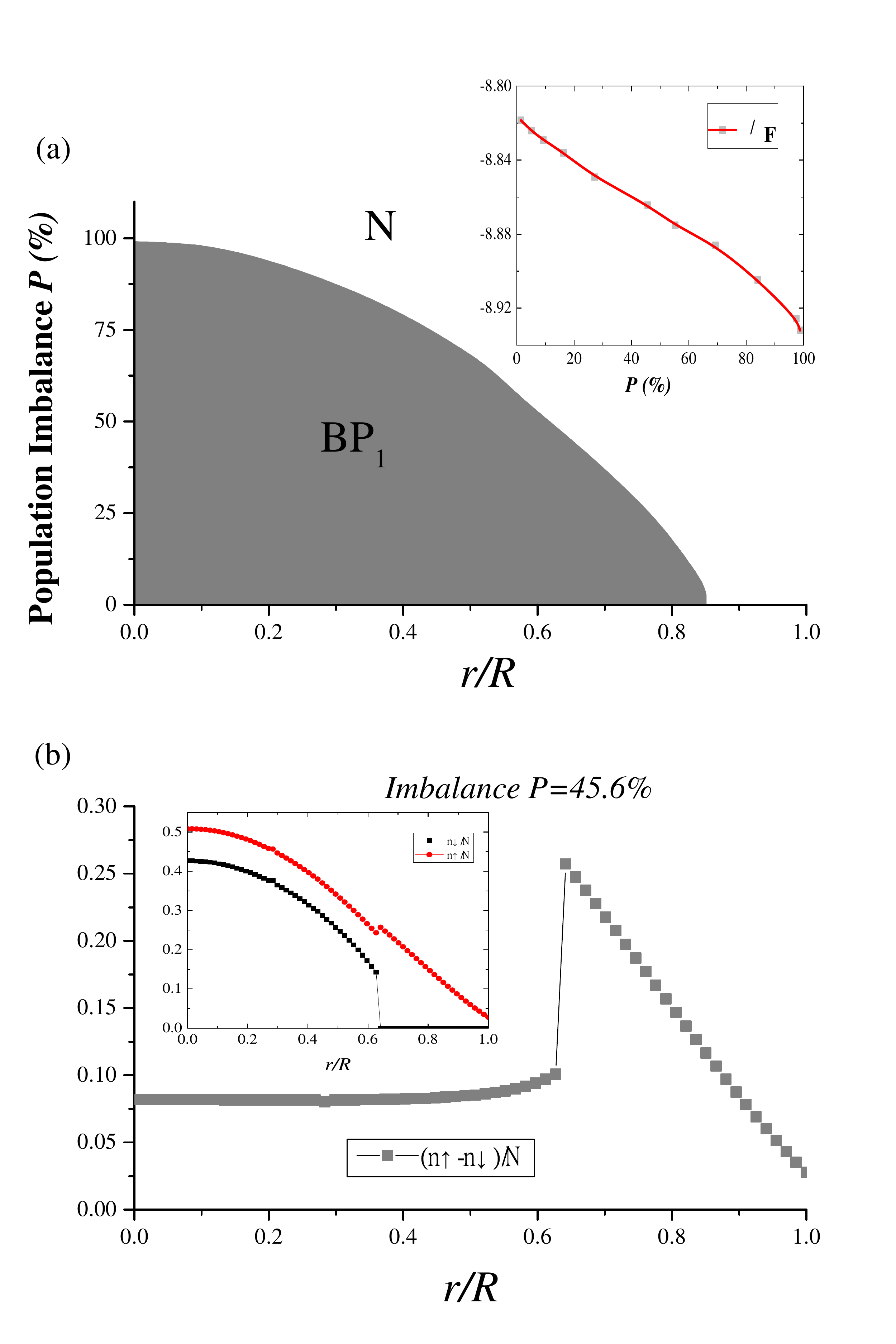}
	\caption{ {\bf  (a)} Phase diagram for the BEC region in the $(r,p)$ plane extracted from the data for the ground state energy density as a function of the distance from the center of the trap. Inset is $\mu$ as function of Imbalance. {\bf  (b)} local spin/population densities and its ratio with local total number. The coupling constant is $\lam=5.95$.}
	\label{Fig6}
\end{figure}

\begin{figure}
	\centering
	\includegraphics[width=7.5cm]{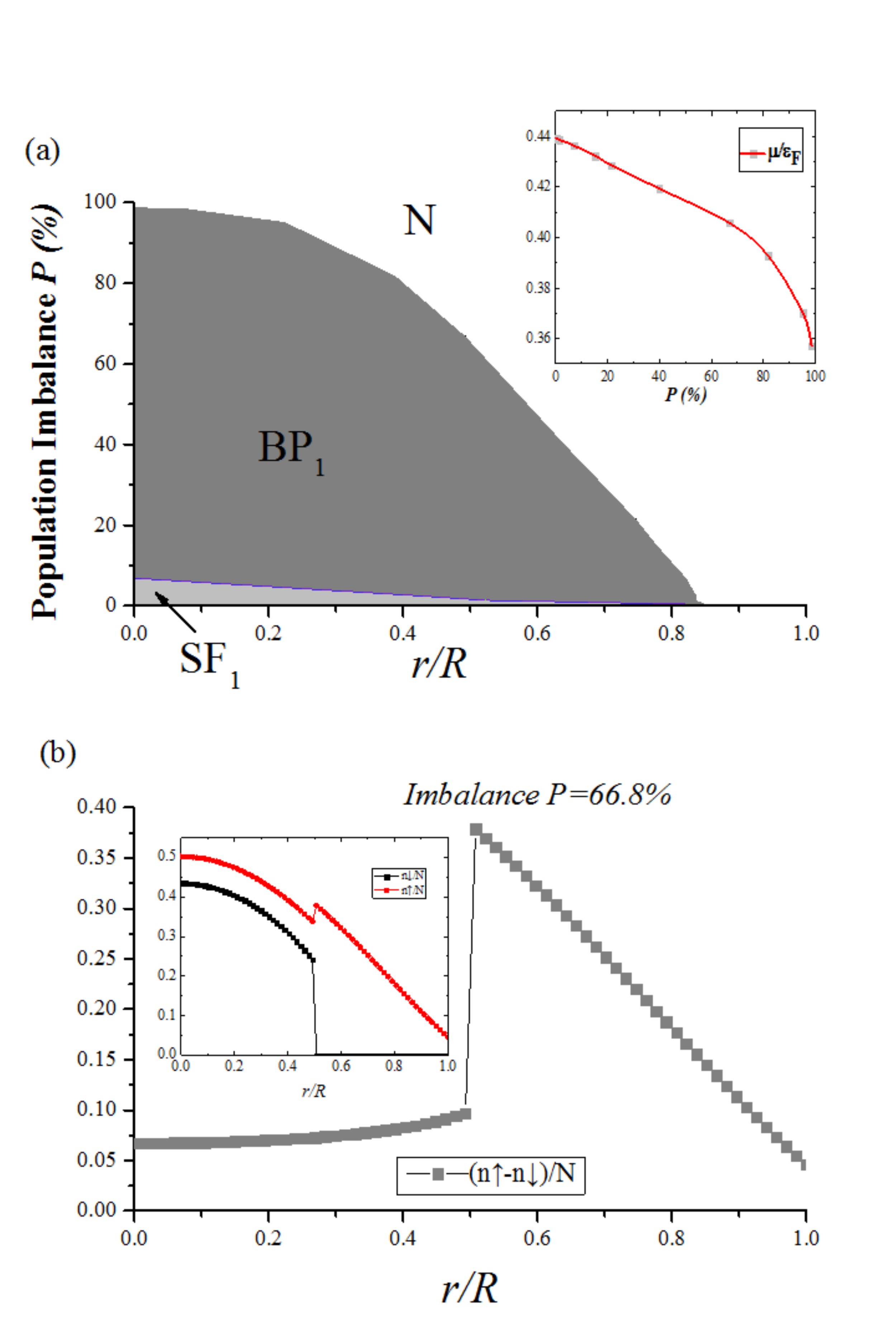}
	\caption{ {\bf  (a)} Phase diagram for the crossover region in the $(r,p)$ plane extracted from the data for the ground state energy density as a function of the distance from the center of the trap. Inset is $\mu$ as function of Imbalance. {\bf  (b)} local spin/population densities and its ratio with local total number. The coupling constant is $\lam=2.9$.}
	\label{Fig7}
\end{figure}

\beg\label{totalNumber1ch}
1=\frac{3}{\pi^3}\int d^3{\tilde{r}} \int d^3\tilde{\bk} \left[1-\frac{\xi_{\tilde{\bk}}}{E_{\tilde{\bk}}}\theta(E_{\tilde{\bk}}-h)\right],
\en
where $\tilde{r}=r/R_{TF}$ , $\tilde{k}=k/k_F$ while the {gap} components and single particle energies are normalized by the Fermi energy$E_F$. Lastly, in the local density approximation the population imbalance $P$ can be computed according to the following equation
\beg\label{imb}
P=\frac{3}{\pi^3}\int d^3{\tilde{r}} \int d^3\tilde{\bk} \theta(h-E_{\tilde{\bk}}),
\en
The {chemical} potential at any distance from trap center is given as $\mu(\tilde{r})=\mu=\mu_0-\tilde{r}^2$ in the units of Fermi energy {$E_F$}. The {free energy} in one-channel model takes the following form
\beg\label{free1ch}
\begin{split}
	E_{\textrm{gs}}=&\sum\limits_{\bk}\left[\xi_\bk-E_\bk+(E_\bk-h)\cdot\theta(h-E_\bk)\right]\\&+
	\sum\limits_{m=-1}^{1}\frac{|\Delta_m|^2}{\lambda},
\end{split}
\en 

We present phase diagrams for both BCS and BEC side of Feshbach Resonance along with the local densities profiles. While solving for $\mu_0$ i.e. the chemical potential at the trap's center, we make sure that we only accept lowest energy state pairing (SF$_1$) 
if its free energy Eq. \ref{free1ch} is less than un-paired state at any distance from trap center. While solving for $\mu_0$ the Ginzburg-Landau expansion dictates that lowest energy pairing {state} is the {orbital ferromagnet.} We then generate phase diagram in (r,p) plane by computing local polarization density. We also present some densities profiles for certain polarization at BEC and BCS side of FR.    

\textit{{{BCS regime.}}} In Fig.~\ref{Fig5} we present phase diagram for BCS side of the FR {based on consideration of} the lowest free energy pairing state  SF$_1$. We {find} that as we increase population {imbalance,} {superfluid} core decreases its radius analogous to s-wave case. The {superfluid} remains unpolarized at the trap center until imbalance is increased to about 18\%. We also show that for finite population imbalance our systems goes from SF/BP$_1$ state to normal state (N) through first order phase transition as can be seen from finite jump in $n_{\up}-n_{\dn}$. {The} {inset in} Fig.~\ref{Fig5}(a) {shows} the dependence of $\mu_0$ (in SF$_1$ state) on the population imbalance.

We define our breached pair state BP$_1$ analogous to two-channel model where $({n_\up({\mathbf r})-n_{\dn}({\mathbf r})})/({n_\up({\mathbf r})+n_{\dn}({\mathbf r})})\not=0$, which also corresponds to the emergence of the gapless excitations in given momentum interval where $\sqrt{\xi_\bk^2+|\Delta(\hat{\bk})|^2}-h\leq 0$. 

\textit{{{BEC regime.}}} As one can see from Fig.~\ref{Fig6}  on the BEC side of FR we {find} that 
{the superfluid}  core exists {only} in form of BP$_1$ and extends to much larger population imbalance compared to BCS side. There is still  the first order phase transition to {the} normal state for high population imbalance as shown in Fig.~\ref{Fig6}. We also {show in the inset} the dependence of chemical potential at trap's center on population imbalance.

\textit{{{Crossover regime.}}}  {In the crossover} regime we present the phase diagram showing 
{essentially the trend} in between BEC and BCS side. We have both SF$_1$ and well as BP$_1$ phases present up to finite population imbalance. As higher imbalances we have only BP$_1$ state analogous to BEC side which is present up to quite high population imbalances. From Fig.~\ref{Fig7} (b), we can see that phase transition between BP$_1$ and normal state is first order in nature for high population imbalances. In the inset in Fig.~\ref{Fig7} (a) we plot the dependence of chemical potential at the trap's center {against} population imbalance.

To summarize, in this section we have shown that under LDA the p-wave superfluid is energetically favorable and exist as core at the trap's center, analogous to the s-wave case. The superfluid exists in a superfluid {orbital ferromagnet} state (SF$_1$) or breached pair state (BP$_1$). The normal state remains polarized as seen from the particle density plots. Finally, the radius of superfluid core decreases with increasing polarization and at certain polarization superfluid core disappears as it is not energetically favorable anymore.


\section{Conclusions}
In this paper we have studied the problem of the {$m_s=0$ triplet} $p$-wave pairing for an atomic Fermi gas subject to a parabolic 
spherical trapping potential within the mean-field approach. We presented the phase diagrams for both two- and one-channel pairing models. In particular, the mean-field results of the one-channel model are relevant for sufficiently wide resonances,
 and away from the exact unitarity limit.
We found that (i) { as would be expected for a rotationally invariant case,} the trapping potential does not affect the degeneracy between the superfluid state SF$_c$ with all non-zero angular momentum components of ${\vec \Delta}$ and the time-reversal symmetry broken state SF$_1$ when only one component of ${\vec \Delta}$ with either $m=1$ or $m=-1$ is non-zero (so-called orbital ferromagnet); (ii) perhaps a somewhat less expected result is that close to the center of the trap, {the} time-reversal-breaking doubly degenerate superfluid state SF$_1$ which also has the same energy as 
SF$_c$ remains the ground state
for any value of the population imbalance until the normal state becomes more energetically favorable.  

\section{Acknowledgements}
The authors express their gratitude to Prof. Emil Yuzbashyan for bringing this problem to their attention and related discussion. The authors also thank 
Anthony Leggett, Jason Ho and Leo Radzihovsky for useful discussions. The work of A. K. and M. D. was financially supported by the National Science Foundation Grant No. DMR-1506547. M.D. thanks DAAD grant from German Academic Exchange Services for the partial financial support and Karlsruhe Institute of Technology, where part of this work has been completed, for hospitality. 

\begin{appendix}
\section{Properties of the self-consistency equations}
The self-consistency equations (\ref{SelfConsist}) have a fairly complicated structure which significantly complicates their numerical analysis.
With the help of the Ginzburgh-Landau expansion, we were able to demonstrate that only few roots will give the minimum value for the free energy provided that the relation (\ref{phases}) is satisfied. Here, we show that the self-consistency equations have other nontrivial solutions
for which the phase relation (\ref{phases}) holds. Thus, in our subsequent discussion we will only be interested in a superfluid state when all the angular momentum components of the pairing wave function are nonzero.
 
Let us re-write the self-consistency equations in a compact matrix form as follows
\beg\label{matrix}
\Delta_m=\sum\limits_{n=-1}^{1}{\cal K}_{mn}[\Delta]\cdot\Delta_n.
\en
The properties of the spherical components entering into (\ref{SelfConsist}) dictate ${\cal K}_{-1,1}={\cal K}_{1,-1}^*\equiv|{\cal K}_{1}|e^{i\psi}$
and ${\cal K}_{0,1}=-{\cal K}_{-1,0}=|{\cal K}_{01}|e^{i\alpha}$. The diagonal components are purely real and I will use the following notations
${\cal K}_{-1,-1}={\cal K}_{1,1}={\cal K}_d$, ${\cal K}_{00}={\cal K}_0$. As a result, adopting these notations, we cast the system of equations (\ref{SelfConsist}) into the following form:
\beg\label{SelfConsist2}
\begin{split}
1=&{\cal K}_d-|{\cal K}_{01}|\frac{\Delta_0}{|\Delta_{-1}|}e^{i(\alpha-\phi_{-1})}\\&+|{\cal K}_{1}|\frac{|\Delta_{1}|}{|\Delta_{-1}|}e^{i(\psi+\phi_1-\phi_{-1})}, \\
1=&{\cal K}_0-|{\cal K}_{01}|\frac{|\Delta_{-1}|}{\Delta_0}e^{-i(\alpha-\phi_{-1})}\\&+|{\cal K}_{01}|\frac{|\Delta_{1}|}{\Delta_0}e^{i(\alpha+\phi_{1})}, \\
1=&{\cal K}_d+|{\cal K}_{01}|\frac{\Delta_0}{|\Delta_{1}|}e^{-i(\alpha+\phi_1)}\\&
+|{\cal K}_{1}|\frac{|\Delta_{-1}|}{|\Delta_{1}|}e^{-i(\psi+\phi_{1}-\phi_{-1})}
\end{split}
\en
and we took into account that the phases of the components $\Delta_{\pm 1}$ can always be taken relative to the phase of $\Delta_0$, so that the 
latter is considered to be purely real.  We have to keep in mind that all the matrix elements are the functions of $\Delta_0$ and $\Delta_{\pm 1}$. 
Clearly two out of six of these equations are redundant since we have only five unknowns. 

Let us  take an imaginary part from the first and the third equations (\ref{SelfConsist2}) which yields
\beg\label{key}
\begin{split}
&|{\cal K}_{01}|\Delta_0\sin(\alpha-\phi_{-1})=|{\cal K}_{1}|b\sin(\psi+\phi_{1}-\phi_{-1}), \\
&|{\cal K}_{01}|\Delta_0\sin(\alpha+\phi_{1})=-|{\cal K}_{1}|a\sin(\psi+\phi_1-\phi_{-1}).
\end{split}
\en
\textit{In above equation $a,b$ are not formaly defined.}
As we have seen from the Ginzburg-Landau analysis, the minimum of the free energy is achieved when (\ref{phases}) holds implying
$\phi_{1}=-\phi_{-1}$. Then there are two possible scenarios, which we would like to discuss separately.  

\paragraph{Symmetric solution: $|\Delta_{-1}|=|\Delta_{1}|$.}
In the first one, one needs to require $|{\cal K}_{01}|=0$ and $\phi_{-1}=\psi/2$. In turn, equation 
\[
|{\cal K}_{01}(\Delta_{-1},\Delta_{1},\Delta_0)|=0
\]
can only be satisfied for $|\Delta_{-1}|=|\Delta_{1}|$. Clearly, this solution corresponds to the fully condensed superfluid in a state with global minimum of free energy. Furthermore, from the first equation (\ref{SelfConsist2}) it also follows that
\beg\label{phasepsi}
\psi+2\phi_1=\pi n,
\en
where $n$ is an integer, so that the phase of $\Delta_1$ is basically determined by the phase of the off-diagonal matrix element ${\cal K}_{-1,1}$. 
Furthermore, there are two roots corresponding to the even or odd values of $n$ and one needs to compute the free energy to determine which
of two roots correspond to the ground state.

\paragraph{Asymmetric solution: $|\Delta_{-1}|\not=|\Delta_{1}|$.}
In the second scenario we impose the constraint on the non-vanishing $K_{01}$ in (\ref{key}) so that from (\ref{key}) it follows that the following equation must be fulfilled:
\beg\label{A5}
|\Delta_{-1}|\sin(\alpha-\phi_{-1})=-|\Delta_1|\sin(\alpha+\phi_{1}),
\en
which is the imaginary part of the second equation (\ref{SelfConsist2}). Since we still need to search for all possible solutions when 
$\phi_{-1}=-\phi_1$ since this relation guarantees the minimum in 
the free energy as we have seen from the Ginzburgh-Landau analysis. 
Then it immediately follows that equation (\ref{A5}) has the following nontrivial solution
\beg
\alpha-\phi_{-1}=\pi n,
\en
where $n=0,1,2,...$. Furthermore, since in this scenario $|\Delta_{-1}|\not=|\Delta_1|$ from the first two equations we can also obtain the following relation between the order parameter amplitudes:
\beg\label{UsefulRelate}
\Delta_0^2=\left(|\Delta_1|-|\Delta_{-1}|\right)^2\left({\frac{1-{\cal K}_d}{1-{\cal K}_0}}\right).
\en
Thus, as we have seen there are two possible solutions corresponding to the fully condensed state. However, as we have checked by the direct numerical calculation, { the state with $|\Delta_{-1}|\not=|\Delta_1|$ always have higher energy than the state with $|\Delta_{-1}|=|\Delta_1|$ in agreement with the Ginzburg-Landau analysis.} {The state we found such that $|\Delta_1|\neq |\Delta_{-1}|$ was supposed to be with same energy as symmetric state like orbital ferromagnet.}

\end{appendix}

\bibliography{pstrap}

\end{document}